\documentclass{ws-ijmpa}

\newcommand{\secondG}{\mathcal{G}_2}

\newcommand{\fourthG}{\mathcal{G}_4}

\begin{document}

\markboth{G.I. Gomero}
{Can We See the Shape of the Universe?}

\catchline{}{}{}

\title{CAN WE SEE THE SHAPE OF OUR UNIVERSE?\footnote{Talk given at 
the Fif\/th Alexander Friedmann International Seminar on Gravitation 
and Cosmology held in Jo\~ao Pessoa, Para\'{\i}ba, Brazil from April 
24th to April 30th, 2002.}}

\author{\footnotesize G.I. GOMERO\footnote{german@cbpf.br}}

\address{Centro Brasileiro de Pesquisas F\'{\i}sicas, Rua Dr. Xavier 
Sigaud 150 \\
22290-180, Rio de Janeiro -- RJ, Brazil}

\maketitle

\pub{Received (Day Month Year)}{Revised (Day Month Year)}

\begin{abstract}
This is a written version of a talk given at the Fif\/th Friedmann 
Seminar on recent work in Observational Cosmic Topology done in 
partial collaboration with Armando Bernui. We address three relevant 
questions related to the search for the size and shape of our 
Universe: (i) How do the actual observation of multiple images of 
certain cosmic objects, e.g. galaxy clusters, constrain the possible 
models for the shape of our Universe?, (ii) What kind of predictions 
can be done once a pair of cosmic objects have been identif\/ied to 
be topological images related by a translation?, and (iii) Is it 
possible to determine if two regions of space are topologically 
identif\/ied, even when distortions on the distributions of cosmic 
sources due to observational limitations are not negligible? We 
give examples answering the f\/irst two questions using the 
suggestion of Roukema and Edge that the clusters RXJ 1347.5-1145 and 
CL 09104+4109 might be topological images of the Coma cluster. For 
the third question, we suggest a method based on the analysis of 
PSH's noise correlations which seems to give a positive answer.

\end{abstract}

\section{Introduction}

The talk given at the Fif\/th Friedmann Seminar was a brief report 
of recent work in Cosmic Topology done in partial collaboration 
with Armando Bernui (see Refs.~1--3 for the original papers). This 
written version brief\/ly summarizes the main results and discusses 
further work.

What is the shape of our Universe? This is one of the most 
intriguing questions in observational cosmology, and the last two 
decades have seen a continuously increasing interest, from the part 
of cosmologists, in studying problems closely related to it 
(see Ref.~4 and references therein). Active research is 
currently being done from the more formal aspects such as the 
(im)posibility  of observing (at least part of) the shape of 
space,\cite{Detect} up to developing concrete methods to detect 
(or even determine) a non-trivial topology of our Universe. These 
methods are all based on the observation of multiple images of 
discrete cosmic objects\cite{CosCris} or some kind of pattern 
repetition on maps of CMBR.\cite{CMBR}

Three relevant questions related to the search for the size and 
shape of our Universe are: (i) How do the actual observation of 
multiple images of certain cosmic objects, e.g. galaxy clusters, 
might constrain the possible models for the shape of our Universe?, 
(ii) What kind of predictions can be done once a pair of cosmic 
objects have been identif\/ied to be topological images related by 
a translation?, and (iii) Is it possible to determine if two 
regions of space are topologically identif\/ied even when 
distortions on the distributions of cosmic sources due to 
observational limitations are not negligible? 

In Refs.~1 and 2 I gave examples answering the f\/irst two questions 
using a suggestion of Roukema and Edge that the clusters RXJ 
1347.5-1145 and CL 09104+4109 might be topological images of the Coma 
cluster,\cite{RE97} these results are reviewed in Sections \ref{Model} 
and \ref{Images} respectively. For the third question, in Ref.~3 A. 
Bernui and I suggested a method based on the analysis of PSH's noise 
correlations which seems to give a positive answer. This method is 
brief\/ly described in Section \ref{Local}. Finally, Section 
\ref{Conclusion} deals with conclusions and further work.

\section{Model Building in Cosmic Topology}
\label{Model}

Suppose we have identif\/ied three clusters of galaxies as being 
dif\/ferent topological images of the same object. How do these 
multiple images constrain the possible models for the shape of our 
Universe? Some time ago Roukema and Edge suggested that the clusters 
RXJ 1347.5-1145 and CL 09104+4109 might be topological images of the 
Coma cluster.\cite{RE97} The distances of these clusters to Coma 
being 970 and 960$h^{-1}$ $Mpc$ respectively (taking $\Omega_0=1$ 
and $\Lambda=0$), and the angle between them, with the Coma cluster 
at the vertex, being $\approx \! 88^o$. Roukema and Edge suggested 
that the clusters RXJ 1347.5-1145 and CL 09104+4109 might be 
images of Coma by equally spaced pure translations and assumed 
a right angle between the lines joining each high redshift cluster 
with Coma. With these considerations they constructed FL 
cosmological models with compact f\/lat spatial sections of constant 
time. These spatial sections were (i) 3-torii, (ii) manifolds of 
class $\secondG$, or (iii) manifolds of class $\fourthG$, all with 
square cross sections, and scale along the third direction larger 
than the depth of the catalogue of X-ray clusters used in the 
analysis.%
\footnote{As in previous papers dealing with Euclidean manifolds, 
the notation used here is that of Wolf.\cite{Wolf}}

In Ref.~1 I analyzed the Roukema-Edge hypothesis in the context of 
Friedmann-Lema\^{\i}tre cosmological models with f\/lat spatial 
sections whose matter components are pressureless dust and a 
cosmological constant, and studied the possibility that at least 
one of these high redshift clusters is an image of Coma by a skrew 
motion. In this study it was not assumed that the distances from 
Coma to both of the high redshift clusters are equal, nor that they 
form a right angle (with Coma at the vertex). It was shown that 
this conf\/iguration of clusters can be accomodated within any of 
the six classes of compact orientable 3-dimensional f\/lat space 
forms, providing a plethora of models for the shape of our Universe. 
Furthermore, it was also shown that the identif\/ication of two more 
triples of multiple images of clusters of galaxies, in the 
neighbourhood of the f\/irst one is enough to completely determine 
the topology of space in most of the models proposed. 

Although it could also be considered the possibility that one pair 
of clusters are identif\/ied by a glide ref\/lection, thus giving 
rise to non-orientable manifolds for models for the shape of space, 
this was not done since these cases do not give qualitatively 
dif\/ferent results, and the corresponding calculations can be done 
whenever needed.

\section{Possible Topological Images of Nearby Clusters of Galaxies}
\label{Images}

A common feature of most of the models constructed in Ref.~1 is 
that one of the topological images is related to Coma by a pure 
translation. This gives the possibility of making precise predictions 
for the positions of images of other clusters of galaxies. In 
Ref.~2 it was reported the angular positions and redshifts of these 
potential images for 31 nearby clusters of galaxies from the 
Abell-ACO catalogue.\cite{ACO} Remarkably, several of the predicted 
angular positions coincide with angular positions of faint clusters 
of the same catalogue, within a $1^{\mbox{o}}$ uncertainty, suggesting 
that topological images of clusters might have been detected and 
recorded a long time ago, although they have not been recognized as 
being so. These faint clusters of galaxies have no measured redshifts, 
so it was argued to be of considerable importance to measure them in 
order to verify if they are actually topological images of nearby 
clusters. Moreover, it was suggested to be also of considerable 
interest to plan and execute deep redshift surveys looking for 
clusters of galaxies around the clusters RXJ 1347.5-1145 and CL 
09104+4109, and their antipodal points with respect to Coma, in order 
to identify the (possible) topological images predicted in that work.

\section{Local Correlations in Cosmic Topology}
\label{Local}

Cosmic Crystallography is a statistical method which looks for 
distance correlations between cosmic sources in pair separation 
histograms (PSH), i.e. graphs of the number of pairs of sources 
versus the squared distance between them. These correlations are 
due to isometries, of the covering group of the manifold that 
models the spatial sections of the Universe, which give rise to the 
(observed) multiple images. This method has been extensively studied 
by diverse groups of researchers, although its present stage of 
development does not allow its direct application to real current 
catalogues.\cite{CosCris} 

On the other hand, a method proposed by Roukema in Ref. 11, looks 
for identical quintuplets of quasars in two dif\/ferent regions of 
space. As it stands, this method is not practical since it looks 
for identical rigid conf\/igurations, and does not consider the 
possibility that in one of the conf\/igurations, one or more of the 
quasars are not actually being observed. Nevertheless, an abstraction 
of the 3-D quasars positions proposal lead us to the idea of looking 
at correlations between the distributions of sources in dif\/ferent 
regions of space.

Indeed, in Ref.~3 A. Bernui and I proposed a new method for the 
search of topology that captures ideas from Cosmic Crystallography 
and the 3-D quasars positions method of Roukema. The basic idea is 
to analyse the correlations of the noise of two PSHs corresponding 
to two small distant regions of space. The Local Noise Correlations 
method (or LNC for short), as it was called, calculates the 
coef\/f\/icient of linear correlation, $r$, of the noise of two PSHs. 
If the two regions under scrutiny are not related by an isometry, 
then the correlation coef\/f\/icient vanishes. On the other hand, if 
the two regions are identif\/ied by an isometry, they should have 
ideally an identical distribution of images, so the (ideal) 
coef\/f\/icient of linear correlation takes the value $r=1$. 

Actually, even if the two regions are identif\/ied by an isometry, 
the distributions of images in both regions are not identical. This 
is due mainly to the fact that the two regions are at dif\/ferent 
distances to the observer, and so correspond to dif\/ferent epochs 
of the Universe. As a consequence, peculiar velocities of the 
sources, luminosity thresholds and f\/inite lifetimes of the sources 
contribute to diferentiate the two samples of images with which one 
produces the PSHs. However, and this is very remarkable, the noises 
of both PSHs remain strongly correlated provided the distortions in 
the distributions due to observational limitations are not too 
strong.

The distributions of images in both samples may be non-identical due 
to another reason. The two regions may not be identif\/ied by an 
isometry, but the image of one of the regions may overlap the other 
one, thus some \emph{portions} of both distributions would actually 
coincide. We showed that the LNC method is also robust under this 
source of distortion provided the overlapping region is not too 
small.

Thus, the robustness of the LNC method suggests its potential value 
for the search of the topology of the Universe in at least two 
dif\/ferent ways. First, it was argued that the Roukema-Edge 
hypothesis can be tested in a def\/initive way by the LNC method 
provided detailed redshift surveys can be performed in the regions 
around these two high redshift clusters. In fact, the correlation 
coef\/f\/icient of the noise of the PSHs corresponding to these two 
surveys, together with the same analysis applied to each of these 
surveys with a survey of local clusters of galaxies would tell if 
these three regions are actually identif\/ied by isometries.

The second suggestion is more speculative. It was shown in Ref.~1 
that the identif\/ication of three neighboring triples of multiple 
images of clusters of galaxies is enough to completely determine 
the topology of space in most of the models proposed there. For the 
remaining possibilities one can determine just two generators of 
the covering group and either the axis of rotation of the third 
generator, or at least its direction. If it turns out that the 
clusters RXJ 1347.5-1145 and CL 09104+4109 are actually topological 
images of Coma, then it would be very easy to identify many more 
triples of images of clusters. If, additionally, it turns out that 
the shape of space is such that it cannot be determined by these 
additional clusters, one can use the LNC method to search for the 
remaining generator by comparing regions along the axis of rotation, 
or parallel to it.

\section{Discussion and Further Remarks}  
\label{Conclusion}

In this is written version of my talk given at the Fif\/th 
Friedmann Seminar, I addressed three relevant questions related to 
the search for the size and shape of our Universe. These questions 
deal respectively with model building, the prediction power of these 
models, and the possibility of developing search methods of topology 
which would test them. I reported examples answering the f\/irst two 
questions using the suggestion of Roukema and Edge that the clusters 
RXJ 1347.5-1145 and CL 09104+4109 might be topological images of the 
Coma cluster. As for the third question, I reported a method based 
on the analysis of PSH's noise correlations which seems to be 
of great potential value.

The results reviewed here suggest the importance of developing 
suitable methods for deciding whether two clusters of galaxies are 
topological images. Once three clusters had been suggested as being 
images of the same cluster, the methods discussed here can be used 
to test the hypothesis. In this respect, it turns out to be of 
considerable importance to plan and execute deep redshift surveys 
looking for clusters of galaxies around the clusters RXJ 1347.5-1145 
and CL 09104+4109, and their antipodal points with respect to Coma, 
in order to identify the topological images that may exist if the 
Roukema-Edge hypothesis turns out to be true.

Despite the physical and philosophical relevance of identifying the 
shape of our Universe, there are other related topics of practical 
interest in cosmology and astrophysics. In fact, it may be possible 
to use the knowledge of the shape of our Universe to improve the 
determination of cosmological parameters\cite{Curv} and to measure 
transverse velocities and radial positions of galaxies in 
clusters.\cite{GalaxVel} Although these studies are in their very 
begining, one can expect that further developments would open the 
possibility of considerably improve our knowledge of the time 
evolution of galaxies and of cluster dynamics. It is then of 
considerable interest to determine whether our Universe allows the 
existence of multiple images of these objects.

\vspace{3mm}
\section*{Acknowledgements}
I would like to thank CLAF for the grant under which this work 
was carried out.


%
%
\end{document}